\documentclass [a4paper,12pt]{article} 

\usepackage{amsmath, amsfonts, amssymb, geometry, fancyhdr, mathrsfs}
\usepackage{fancyhdr, dsfont}
\usepackage[latin1]{inputenc}
\usepackage[pdftex]{graphicx}
\usepackage{float}
\usepackage{color}
\definecolor{dgreen}{rgb}{0,0.70,0.30}
\definecolor{gold}{rgb}{0.85,.66,0}

\advance\voffset by -1.5cm
\advance\hoffset by -0.50cm
\catcode`\ä = \active
\catcode`\Ä = \active
\catcode`\ö = \active
\catcode`\Ö = \active
\catcode`\ü = \active
\catcode`\Ü = \active
\catcode`\ß = \active
\defÄ{"A}
\defä{"a}
\defÖ{"O}
\defö{"o}
\defÜ{"U}
\defü{"u}
\defß{"s}
\def\beq{\begin{equation}}
\def\eeq{\end{equation}}

\newcommand{\vecb}{\left(\begin{array}{c}}
\newcommand{\vece}{\end{array}\right)}
\newcommand{\ccb}{\left(\begin{array}{cc}}
\newcommand{\cce}{\end{array}\right)}
\newcommand{\cccb}{\left(\begin{array}{ccc}}
\newcommand{\ccce}{\end{array}\right)}
\newcommand{\ccccb}{\left(\begin{array}{cccc}}
\newcommand{\cccce}{\end{array}\right)}
\newcommand{\cccccb}{\left(\begin{array}{ccccc}}
\newcommand{\ccccce}{\end{array}\right)}

\newcommand{\pa}{\partial}

\newcommand{\al}{\alpha}

\newcommand{\de}{\delta}

\newcommand{\vep}{\varepsilon}

\newcommand{\si}{\sigma}
\newcommand{\la}{\lambda}

\newcommand{\Ga}{\Gamma}

\newcommand{\te}{\textrm}
\newcommand{\eq}{ \ = \ }

\newcommand{\co}{\ , \ \ \ \ \ \ }

\newcommand{\dd}{\mathrm{d}}

\newcommand{\ZZ}{\mathbb Z}

\def\beq{\begin{equation}}
\def\eeq{\end{equation}}

\begin{document}

\vspace*{1.5cm}

\begin{center}
{\Large 
{\bf Bulk induced boundary perturbations} 
\vspace{0.5cm}

{\bf  for ${\cal N}=1$ superconformal field theories}}
\vspace{2.5cm}

{\large Matthias R.\ Gaberdiel$^{1}$}
\footnotetext{$^{1}${\tt E-mail: gaberdiel@itp.phys.ethz.ch}} 

Institut f{\"u}r Theoretische Physik, ETH Z{\"u}rich\\
CH-8093 Z{\"u}rich, Switzerland\\
\vspace*{0.5cm}

{\large and}  
\vspace*{0.5cm}

{\large Oliver Schlotterer$^{2}$}

\footnotetext{$^{2}${\tt E-mail: oliver.schlotterer@web.de}}

Max-Planck-Institut f{\"u}r Physik (Werner-Heisenberg-Institut) \\
D-80805 M{\"u}nchen, Germany

\vspace*{3cm}

{\bf Abstract}
\end{center}
The ${\cal N}=1$ superconformal circle theory
consisting of a free boson and a free fermion is considered.
At any radius the theory has standard Dirichlet and Neumann
branes, but for rational radii there are additional superconformal 
boundary conditions that are labelled by elements in a quotient of
SU(2). We analyse how these branes behave under the 
radius-changing bulk perturbation.
As in the bosonic case, the bulk perturbation induces in general
a boundary RG flow whose end-point is a superposition of
Dirichlet or Neumann branes.


\newpage
\renewcommand{\theequation}{\arabic{section}.\arabic{equation}}


\section{Introduction}
\setcounter{equation}{0}

Many closed string backgrounds possess moduli that allow one to change
the shape and size of the background geometry. Furthermore, the
possible D-branes of a given background also typically form a moduli
space \cite{Recknagel:1998ih}. Obviously, these two moduli spaces are
not unrelated: the moduli space of D-branes typically depends on 
where one sits in the closed string moduli space, and conversely,
D-branes backreact on the geometry and may have an impact on 
whether some of the closed string moduli may get
lifted. It is clearly an important question to understand in some detail how 
these two moduli spaces are related to one another. 

Recently, some progress has been made by
studying this question from a conformal field theory point of view.
As was shown in \cite{PERT,Fredenhagen:2007rx}, an exactly marginal bulk operator
(describing a bulk deformation) can cease to be
exactly marginal in the presence of a boundary. If this is the case,
it induces a non-trivial RG flow on the boundary that drives the
boundary condition to one that is compatible with the deformed closed
string background. 

As an example, this process was studied for the case of a single free
boson in \cite{PERT}. The moduli space of D-branes for this theory depends 
crucially on the radius of the circle \cite{GA,Janik,Friedan}: 
for all radii there are Neumann and Dirichlet 
branes, but if the radius is a rational multiple of the self-dual radius,
there is an additional 3-dimensional branch of the moduli space of 
conformal D-branes. On the other hand, for irrational multiples of the
self-dual radius, the additional branch of the moduli space 
is only 1-dimensional. The structure
of the full moduli space of conformal D-branes thus changes very 
discontinuously as one varies the radius of the circle theory. In fact,
if one starts with a generic brane 
at a rational point, then the radius-changing bulk 
perturbation is not exactly marginal, but rather induces a non-trivial RG flow on the boundary.
In the example at hand this RG flow could be solved exactly
(using the equivalence of the theory at the self-dual radius to the SU(2) WZW
model at $k=1$), and the end-point of the flow could be determined \cite{PERT}: if 
the radius is increased, a generic brane always flows to a (superposition of)
Dirichlet branes, while if the radius is decreased, the endpoint of
the flow is a (superposition of) Neumann branes.
\smallskip

In this paper we study the ${\cal N}=1$ supersymmetric analogue of this problem. The
moduli space of ${\cal N}=1$ superconformal branes for the free boson
and fermion theory has a similar structure as in the bosonic case 
\cite{GA,Sen1},  and there is also a close relation to the SU(2) WZW model, 
this time at $k=2$ \cite{DGH,MP}.\footnote{The rational boundary states 
for all multicritical points were also constructed in \cite{Cappelli:2002wq}.}
However, there are also some differences: the
WZW model description only applies to the superaffine theory at $R=1$ (not one of 
the circle theories), and one needs to keep track carefully of the GSO-projection. 
As we shall show, one can overcome these difficulties and obtain as 
complete a picture as in the bosonic example. In particular, one finds
that generic branes flow to superpositions of Dirichlet or Neumann
branes as the radius of the circle is increased or decreased, respectively.

Bulk induced boundary perturbations have also been discussed in 
\cite{Green:2006ku,Gaberdiel:2007us,Baumgartl:2007an,Green:2007wr}, 
as well as in the context of defect operators
\cite{Brunner1,Brunner:2007ur}. The backreaction
effect has also been analysed from this point of view in \cite{Keller:2007nd}.
\medskip

The paper is organised as follows. In section~2 we briefly review the 
salient features of the bosonic analysis. Section~3 explains how the
${\cal N}=1$ free boson and free fermion theory is related to the
WZW model at $k=2$, and the corresponding boundary states  are 
identified in section~4. In section~5 we then put everything together
and deduce the RG flow of the ${\cal N}=1$ superconformal branes
from the WZW analysis. Section~6 contains our conclusions. There is one
appendix containing some technical calculation.

\section{Review of the bosonic analysis}
\label{sec:bosonic}

Let us begin by reviewing briefly the analysis in the bosonic case. We 
consider the $c=1$  conformal field theory of a single free boson 
compactified at a radius $R$. At the self-dual radius $R=1/\sqrt{2}$ 
--- in our conventions $\alpha'=\tfrac{1}{2}$ --- the theory is equivalent to 
the SU(2) WZW model at level $1$, where the left-moving currents are expressed
in terms of the left-moving free boson field $X_L$ as
\beq
J^{3}(z) \ := \ i\sqrt{2} \, \pa_{z} X_{L}(z) \co 
J^{\pm}(z) \ := \ : \te{exp} \bigl(\pm 2\sqrt{2}i \, X_{L}(z) \bigr) : \ ,
\label{1,A}
\eeq
and similarly for the right-movers such as 
$\bar{J}^{3} = -i\sqrt{2} \pa_{\bar{z}} X_{R}$. 
It was shown in \cite{SU2} that 
the full moduli space of conformal boundary conditions for this theory 
is precisely the group manifold SU(2). The corresponding boundary conditions
preserve the su(2) affine symmetry up to  conjugation by a group element;
we choose the conventions that the boundary condition labelled by $g$
satisfies the gluing condition
\beq
\Bigl(\te{Ad}_{(g\cdot \iota)}(J^{a}_{n}) \ + \ \bar{J}^{a}_{-n} \Bigr) \, 
| \! | g\rangle \! \rangle_{\te{WZW}} \eq 0 \ ,
\label{1,B}
\eeq
where $\iota$ is the SU(2) matrix 
$\iota := \left( \begin{smallmatrix} 0 &1 \\ -1 &0 \end{smallmatrix} \right)$, and
Ad denotes the adjoint representation of SU(2).
In these conventions, a diagonal group element describes a Dirichlet brane, 
while off-diagonal group elements correspond to Neumann branes.
\smallskip

If the radius of the circle is a rational multiple of the self-dual radius,
$R=\tfrac{M}{N} \tfrac{1}{\sqrt{2}}$, then the conformal boundary conditions
are labelled by elements in the quotient space 
SU(2)$/{\mathbb Z}_M\times {\mathbb Z}_N$. 
In addition, there are also the usual Dirichlet and Neumann branes (that exist
at any radius). On the other hand, if the radius is an {\it irrational} multiple of the 
self-dual radius, the moduli space of boundary conditions is much smaller 
\cite{GA,Janik,Friedan}.  
\smallskip

In the bulk theory, the radius $R$ is a modulus of the theory, {\it i.e.}\ it corresponds
to an exactly marginal bulk operator $\Phi$.  Changing the radius then corresponds
to the perturbation of the action by
\beq
{\cal S}_{\de R} \ = \ {\la} \int \dd^{2}z \ \Phi(z,\bar{z}) \\ = 
 \ \ {\la} \int \dd^{2}z \ J^{3}(z) \, \bar{J}^{3}(\bar{z}) 
\eq  2 \, {\la} \int \dd^{2}z \ \pa_{z}X \, \pa_{\bar{z}}X \ ,
\label{1,D}
\eeq
where in our conventions $\lambda>0$ means that the radius $R$ is decreased. 
As we have mentioned above, the moduli space of D-branes depends in a very
discontinuous manner on the radius, and thus the effect of this bulk perturbation on 
the boundary conditions must be non-trivial. This question was studied in  
\cite{PERT}, where it was shown that the above bulk perturbation is not 
necessarily exactly marginal in the presence of a boundary, but rather induces 
in general a non-trivial RG flow on the boundary. If we denote the boundary 
coupling constants by $\mu_k$, then the relevant RG equations are of the form
\beq
\dot\mu_k  \eq (1-h_{k}) \, \mu_k \ + \ \frac{1}{2}\, B_{\Phi k}\, \lambda \
+ \ D_{ijk}\, \mu_i \, \mu_j \  + \ {\cal O}(\mu\lambda, \mu^3, \lambda^2)\ .
\eeq
Here $h_k$ is the conformal dimension of the boundary field corresponding to 
$\mu_k$ --- in our case the boundary fields of interest are just the currents whose
conformal dimension is equal to $1$ --- while $B_{\Phi k}$ denotes 
the bulk-boundary coupling constant of the perturbing bulk field $\Phi$, and
$D_{ijk}$ are the boundary OPE coefficients.
For the case at hand, the bulk boundary coefficient could be determined
explicitly as\footnote{Because the boundary conditions are labelled with
the inclusion of $\iota$, there is now a different sign compared to \cite{PERT}. 
Note however that $\lambda>0$ now corresponds to decreasing the radius.} 
\beq\label{bulkboundary}
B_{\Phi \gamma} \ = \   i \, \hbox{Tr} \Bigl( t^\gamma \, [t^3, g t^3 g^{-1} ] \Bigr)  \ ,
\eeq
where $t^\alpha$ are the generators of the Lie algebra su(2), and $\gamma$
labels the boundary current $J^\gamma$, while the group element $g$ characterises the
boundary condition in question. To first order in the bulk perturbation, 
the induced boundary RG flow only changes the group element $g$, and it 
is then possible to integrate up the induced boundary flow completely (to first
order in $\lambda$). If we label the group elements in SU(2) as 
\beq
g \eq \ccb e^{i\phi} \, \cos \theta &i e^{i\psi} \, \sin \theta \\ 
i e^{-i\psi} \, \sin \theta &e^{-i\phi} \, \cos \theta \cce \ ,
\label{1,F}
\eeq
then the brane flow only affects $\theta$; if the radius is increased (decreased)
the brane labelled by $g$ flows to a pure Dirichlet (Neumann) brane whose
value of $\phi$ ($\psi$) is unchanged.

\section{Equivalence of bulk theories}
\label{sec:susybulk}
\setcounter{equation}{0}

In the following we want to repeat this analysis for the ${\cal N}=1$ superconformal
field theory consisting of a free boson and a free fermion. As we have seen above,
the bosonic analysis was simplest in the WZW description of the theory. In the superconformal
case, there is also a direct relation to a WZW model --- this time the su(2) WZW
model at level $2$  --- and it will again be convenient to use this formulation of the
theory. In the superconformal context the WZW model is not directly equivalent 
to the theory of a free boson and fermion, but rather to the so-called
superaffine theory \cite{DGH}. 
\smallskip

Let us first describe the WZW model in some detail. It is well known that the 
su(2) level $2$ theory has a free field realisation in terms
of three free Majorana fermions, see for example \cite{GO}. Let us denote the 
corresponding fermion fields as $\psi^a(z)$, where $a$ denotes the adjoint
representation of su(2), and we have the commutation relations
\beq
{}\{ \psi^a_r,\psi^b_s\} \eq \delta_{r,-s}\, \delta^{ab} \ . 
\eeq
Then the WZW currents are given as 
\beq
J^{a}(z) \ = \ -\frac{i}{2} \; \vep^{abc} \, : \psi^{b}(z) \, \psi^{c}(z) : 
 \ \ \ \Longleftrightarrow \ \ \
J^{a}_{n} \eq - \frac{i}{2}
\sum_{r } \vep^{abc} \, : \psi^{b}_{n-r} \, \psi^{c}_{r} : \ ,
\label{2,14}
\eeq
where $\vep^{abc}$ is the totally antisymmetric tensor in three
dimensions. 

At level $2$, the possible representations of the su(2) affine algebra 
are ${\cal H}_j$ with $j=0,\tfrac{1}{2},1$, and the complete space of states is of the form
\beq
{\cal H}_{\te{WZW}} \ = \ \bigl( {\cal H}_{0} \otimes \bar{{\cal H}}_{0} \bigr) 
\ \oplus \ 
\bigl( {\cal H}_{\frac{1}{2}} \otimes \bar{{\cal H}}_{\frac{1}{2}} \bigr) 
\ \oplus \
\bigl( {\cal H}_{1} \otimes \bar{{\cal H}}_{1} \bigr) \ . 
\label{2,11}
\eeq
In terms of the free fermion description, the first and last term arise from the
NS-NS sector, while the middle term (the representation $j=\tfrac{1}{2}$) 
corresponds to the R-R sector. Each of these sectors is moded out by the GSO-projection 
$\tfrac{1}{2}(1+(-1)^{F+\tilde{F}})$; in fact, we have the simple relation between 
the $\hat{su}(2)_{2}$ characters 
\beq
\chi_{j=0}(q) \ = \ \frac{f_3(q)^3  +  f_4(q)^3}{2} \ , \ \ \ \chi_{j=1}(q) \ = \ \frac{f_3(q)^3  -  f_4(q)^3}{2} \ , \ \ \  \chi_{j=\tfrac{1}{2}}(q) \ = \ \frac{1}{\sqrt{2}} f_2(q)^3  \ ,
\label{ffch}
\eeq
where the functions $f_i(q)$ are the usual functions of \cite{Pol} that
describe the characters of free fermion representations, see
(\ref{ffun}). 
\medskip

This WZW model is now equivalent to the so-called superaffine theory that is 
defined as a ${\mathbb Z}_2$ orbifold of the free boson and fermion circle 
theory at radius $R=1$. Let us denote the modes of the boson field $X$ 
of the circle theory by $\alpha_m$ and $\bar\alpha_m$,
while the modes of the free fermion fields $\chi(z)$ and $\bar\chi(\bar{z})$
are denoted by $\chi_r$ and $\bar\chi_r$; the commutation relations are
\beq
{}[\alpha_m,\alpha_n] \eq m \, \delta_{m,-n} \ , \qquad
{}[\alpha_m,\chi_r] \eq 0 \ , \qquad
{}\{\chi_r,\chi_s\} \eq  \delta_{r,-s} \ 
\eeq
and similarly for the right-moving modes. The orbifold acts as 
\beq
{\cal S} \eq \Bigl( X \mapsto X + \pi R\Bigr) \times (-1)^{F_{st}} \ ,
\eeq
and $F_{st}$ denotes the left-moving spacetime fermion number,
such that $(-1)^{F_{st}}$ acts as $+1$ ($-1$) on the NS-NS (R-R) sector. 
In the untwisted sector of this orbifold, the left- and right-moving momenta are
thus of the form 
\beq
\hbox{untwisted:}\quad
(p_{L},p_{R})  \eq \left( \frac{k}{2} \, + \, w \ , \ \frac{k}{2} \, - \, w \right) \co 
\left\{ \begin{array}{rl} k \in 2\ZZ, \ w \in \ZZ &: \ \te{NSNS} \\ k \in 2\ZZ-1, 
\ w \in \ZZ &: \ \te{RR ,} \end{array} \right.
\label{2,2}
\eeq
while in the twisted sector we have instead 
\beq
\hbox{twisted:}\quad
(p_{L},p_{R}) \eq \left( \frac{k}{2} \, + \, w \ , \ \frac{k}{2} \, - \, w \right) \co 
\left\{ \begin{array}{rl} k \in 2\ZZ-1, \ w \in \ZZ-\frac{1}{2} &: \ \te{NSNS} \\ 
k \in 2\ZZ, \ w \in \ZZ-\frac{1}{2} &: \ \te{RR .} \end{array} \right.
\label{2,3}
\eeq
In the twisted sector the GSO-projection is reversed; thus in the twisted NS-NS
sector the momentum ground state is now odd under the GSO-projection.

It is instructive to understand how the currents of the WZW description appear
in the superaffine orbifold. First we observe that the currents $\partial X$ and
$\bar\partial X$ are invariant under the orbifold projection; they correspond to
the two currents $J^3$ and $\bar{J}^3$ in the WZW language. The other
currents of the WZW theory arise as the fermionic descendants of the 
momentum ground states of the twisted sector
\beq
J^\pm  \ \ \ \Longleftrightarrow \ \ \ \chi_{-1/2} \, |(p_{L}=\pm 1, p_{R}=0)\rangle \ , \qquad
\bar{J}^\pm \ \ \ \Longleftrightarrow \ \ \ \bar\chi_{-1/2} \, |(p_{L}=0,p_{R}=\pm 1)\rangle \ .
\eeq
It is also fairly straightforward to show that the partition function of 
the WZW model agrees with that of the superaffine theory. This is
most easily seen by writing the WZW model partition function
in terms of free fermion characters
\beq
Z_{\te{WZW}} (q,\bar{q}) \eq \frac{1}{2} \Bigl( |f_3(q)|^6 \, + \, |f_4(q)|^6 \, + \, |f_2(q)|^6 \Bigr)\ , 
\label{Z}
\eeq
as follows from (\ref{ffch}). On the other hand, using the sum representations
of the theta functions, one can show that (\ref{Z}) agrees with the 
partition function coming from the momentum lattice 
(\ref{2,2}) and (\ref{2,3}).

Finally, we note that we can also obtain the circle theory from the superaffine
theory by doing the `quantum orbifold'. In the present case this is the winding shift
orbifold 
\beq\label{quo}
\tilde{\cal S} \eq \Bigl( \tilde{X} \mapsto \tilde{X} + \frac{\pi}{R} \Bigr) \ ,
\eeq
where $\tilde{X}$ is the dual coordinate, {\it i.e.}\ $\tilde{X} = X_L - X_R$,
see also \cite{MP}. 

\section{Boundary conditions}
\label{sec:BoundaryConditions}
\setcounter{equation}{0}

In order to analyse the bulk induced boundary flow for the 
${\cal N}=1$ circle theory, we shall proceed in two steps. We shall
first analyse the situation for the superaffine theory (which is 
equivalent to the WZW model at level $2$), and then deduce from this the
results for the circle theory at radius $R=1$ by considering the circle
theory as the ${\mathbb Z}_2$ orbifold of the superaffine theory. In
order to translate between the different descriptions, we first
need to understand the dictionary between the 
brane descriptions in the different setups in some detail; some
aspects of this were already analysed in \cite{MP}.

\subsection{Branes in the superaffine theory}

We first review the description of the superaffine branes from \cite{MP}. 
We are interested in the branes $|\!| B\rangle\!\rangle$ that preserve the superconformal
(but not necessarily any larger) symmetry. The corresponding gluing conditions 
read
\beq
\Bigl( L_n \ - \ \bar{L}_{-n} \Bigr) \, |\!| B\rangle\!\rangle \eq 0 \eq 
\Bigl( G_r \ + \ i \eta \, \bar{G}_{-r} \Bigr) \, |\!| B\rangle\!\rangle  \ ,
\eeq
where $G$ (and $\bar{G}$) denotes the supercurrent of the ${\cal N}=1$
superconformal algebra --- we use the same conventions as in 
\cite{GA} --- and $\eta=\pm$ labels the two possible choices for 
the ${\cal N}=1$ gluing conditions. 

As explained in \cite{MP}, the Ishibashi states of the superaffine theory
at $R=1$ are labelled by
the usual triplets $(j;m,n)$, together with the sign $\eta=\pm$. In the 
NS-NS sector all combinations with $j$ integer appear, while $j$ is 
half-integer in the R-R sector. We also choose the convention
(as in \cite{MP}) that the R-R sector Ishibashi states are only 
GSO-invariant  for $\eta=-$. With these preparations we can
then give an explicit formula for the boundary states of the superaffine
theory. Depending on the choice of $\eta$ we have 
\begin{align}
| \! | g; - \rangle \! \rangle_{sa} \ &= \ \frac{1}{\sqrt{2}} 
\left( \sum_{(j,m,n) \in {\cal I}^{\te{NS}}} \! \! \! \! \! \! 
D^{j}_{m,n}(g) \, |j;m,n; - \rangle \! \rangle^{\te{NS}} \ 
+ \ \sum_{(j,m,n) \in {\cal I}^{\te{R}}} \! \! \! \! \! \! 
D^{j}_{m,n}(g) \, |j;m,n; - \rangle \! \rangle^{\te{R}} \right) \notag \\
| \! | {g}; + \rangle \! \rangle_{sa} \ &= \ 
\sum_{(j,m,n) \in {\cal I}^{\te{NS}}} \! \! \! \! \! \! 
D^{j}_{m,n}(g ) \, |j;m,n; + \rangle \! \rangle^{\te{NS}} \ .
\label{3,1}
\end{align}
Here ${\cal I}^{\te{NS}}$ contains all the integer spin triplets $(j;m,n)$,
whereas ${\cal I}^{\te{R}}$ covers all the half-integer spin cases.

We shall often refer to the first family of branes ($\eta=-$) as the 
BPS branes, while the second family ($\eta=+$) will be called
non-BPS. It is easy to see that the moduli space of the BPS branes
is precisely SU(2), {\it i.e.} branes corresponding to different group
elements are indeed different. On the other hand, for $\eta=+$, 
the boundary states corresponding to $g$ and $-g$ are identical,
and thus the moduli space is in fact SO(3)=SU(2)$/{\mathbb Z}_2$. 

In general these boundary states only preserve the superconformal
symmetry, but there are special cases that actually preserve more. In 
particular, one easily checks that 
\begin{align}
\bigl( \al_{n} \ - \ \bar{\al}_{-n} \bigr) \, | \! | 
\left( \begin{smallmatrix} e^{i\phi} &0 \\ 0 &
e^{-i\phi} \end{smallmatrix} \right);\eta \rangle \! \rangle_{sa} \ 
&= \ \bigl( \chi_{r} \ + \ i \eta \, \bar{\chi}_{-r} \bigr) \, | \! | 
\left( \begin{smallmatrix} e^{i\phi} &0 \\ 0 
&e^{-i\phi} \end{smallmatrix} \right);\eta \rangle \! \rangle_{sa} \ \ \! \eq 0 \notag \\ 
\bigl( \al_{n} \ + \ \bar{\al}_{-n} \bigr) \, | \! | \left( \begin{smallmatrix} 0 
&ie^{i\psi} \\ ie^{-i\psi} &0 \end{smallmatrix} \right);\eta \rangle \! \rangle_{sa} \ 
&= \ \bigl( \chi_{r} \ - \ i \eta \, \bar{\chi}_{-r} \bigr) \, | \! | \left( \begin{smallmatrix} 0 
&ie^{i\psi} \\ ie^{-i\psi} &0 \end{smallmatrix} \right);\eta \rangle \! \rangle_{sa} \eq 0 \ .
\label{3,2}
\end{align}
The branes associated to off-diagonal group elements are single Neumann
branes, while for diagonal group elements they always describe 
a superposition of two Dirichlet branes at opposite points on the circle;
in the BPS case $(\eta=-$), the two Dirichlet branes are a brane-anti-brane pair,
while in the non-BPS case $(\eta=+)$ the two Dirichlet branes are both
non-BPS branes. From the point of view of the superaffine theory, both 
of these configurations are however fundamental, {\it i.e.}\ cannot 
be resolved into more elementary branes.

As we mentioned before, the superaffine theory is the $\mathbb{Z}_2$ orbifold
of the circle theory, and vice versa. Given the branes of either
theory, we can obtain the branes of the other theory by the usual orbifold
construction. This was explained in detail in \cite{MP}.

\subsection{The WZW description}
\label{sec:TheBoundaryStateDictionary}

As we shall now explain, all of these branes correspond to D-branes
of the WZW model that preserve the affine symmetry up to conjugation,
{\it i.e.}\ that satisfy
\beq
\Bigl( \te{Ad}_{(g\cdot \iota)} (J^{a}_{n}) \ + \  \bar{J}^{a}_{-n} \Bigr) \,  
| \! | g\rangle \! \rangle_{\te{WZW}} \eq 0 \ .
\eeq
It is straightforward to construct the corresponding boundary states
following \cite{Cardy}. The relevant Ishibashi states \cite{Ishibashi} are
simply obtained from the usual  ($g=\iota^{-1}$) Ishibashi states
by the action of $(g\cdot\iota)$; thus we have three families of
Ishibashi states $|g;j \rangle \! \rangle$, coming from the three 
different sectors $j=0,\tfrac{1}{2},1$ of the theory. 
Expressed in terms of the Ishibashi states of the superaffine theory
labelled by $(j,m,n)$, the relation is 
\begin{align}
|g ;0 \rangle \! \rangle \ + \ |g ;1 \rangle \! \rangle \ \ 
&= \  \! \sum_{(j,m,n) \in {\cal I}^{\te{NS}}} \! \! \! \! D^{j}_{m,n}(g) \, 
|j;m,n;- \rangle \! \rangle^{\te{NS}} \notag \\
|g;0 \rangle \! \rangle \ - \ |g ;1 \rangle \! \rangle \ \ 
&= \  \! \sum_{(j,m,n) \in {\cal I}^{\te{NS}}} \! \! \! \! 
D^{j}_{m,n}(g) \, |j;m,n;+ \rangle \! \rangle^{\te{NS}}  \notag \\
|g;\tfrac{1}{2} \rangle \! \rangle \ &= \  
 2^{-\frac{1}{4}}
 \sum_{(j,m,n) \in {\cal I}^{\te{R}}} \! \! \! D^{j}_{m,n}(g) \, 
 |j;m,n;- \rangle \! \rangle^{\te{R}} \ .
\label{3,20}
\end{align}

Given the description of the Ishibashi states, it is then straightforward to 
construct boundary states following \cite{Cardy}. Using the explicit form
of the $S$-matrix for su(2) at level $2$,
\beq
{\cal S}_{j,j'} \eq \left( \begin{matrix}  \frac{1}{2} &
\frac{1}{\sqrt{2}} &\frac{1}{2} \\ \frac{1}{\sqrt{2}} &0 &-\frac{1}{\sqrt{2}} \\ 
\frac{1}{2} &-\frac{1}{\sqrt{2}} &\frac{1}{2} \end{matrix} \right) \ ,
\eeq
the consistent boundary states are 
\begin{align}
| \! | g ;0 \rangle \! \rangle_{\te{WZW}} \ &= \ 
\frac{1}{\sqrt{2}} \; \Bigl( |g;0 \rangle \! \rangle \ 
+ \ |g ;1 \rangle \! \rangle \Bigr) \ 
+ \ \frac{1}{2^{\frac{1}{4}}} \; |g ;\tfrac{1}{2} \rangle \! \rangle 
\eq | \! | g;-\rangle \! \rangle_{sa} 
\notag \\
| \! | g;\tfrac{1}{2} \rangle \! \rangle_{\te{WZW}}\ &= \ \ \ \ \ \ \ \ \,
|g;0 \rangle \! \rangle \ - \ |g;1 \rangle \! \rangle\ \ \ \ \ \ \ \ \ \ \ \ \ \ \ \ \ \ \ \, \! \eq 
| \! |{g};+ \rangle \! \rangle_{sa} \notag \\
| \! | g ;1 \rangle \! \rangle_{\te{WZW}} \ &= \ \frac{1}{\sqrt{2}} \; 
\Bigl( |g ;0 \rangle \! \rangle \ + \ |g;1 \rangle \! \rangle \Bigr) \ 
- \ \frac{1}{2^{\frac{1}{4}}} \; |g ;\tfrac{1}{2} \rangle \! \rangle \eq | \! 
| \! - \! g ;-\rangle \! \rangle_{sa} \ .
\label{3,18}
\end{align}
Since we can express the WZW Ishibashi states in terms of the Ishibashi states
of the superaffine theory (\ref{3,20}), we can then also deduce the identification of the 
WZW model boundary states with those of the superaffine
theory (see  (\ref{3,18})). Note that the first and last line are compatible, since one 
knows  on general grounds (see for example \cite{PERT}) that
\beq
|\!| g ; j \rangle\!\rangle_{\te{WZW}} \eq |\!| -g; \tfrac{k}{2} - j\rangle\!\rangle_{\te{WZW}} \ ,
\eeq
in agreement with the identification in terms of superaffine branes. Applied to
$j=\tfrac{1}{2}$, this observation also implies that the branes in the middle line are
only associated to SO(3)=SU(2)/${\mathbb Z}_2$. 
One can also verify that this identification in agreement with the various
cylinder overlaps; this is explained in more detail in appendix~A.

Finally, we mention in passing that these branes also have a simple description
in terms of the free fermionic description of the WZW model: the relevant
gluing conditions are of the form
\beq
\Bigl( \te{Ad}_{(g\cdot \iota)} (\psi^a_r ) \ + \ i \eta\, \bar\psi^a_{-r} \Bigr) \; 
|\!| g; \eta \rangle\!\rangle_{\psi} \eq 0 \ .
\eeq
However, this will not be important for the rest of our analysis.

\section{The boundary flow}
\label{sec:boundaryflow}
\setcounter{equation}{0}

Now we are ready to analyse the boundary flow in these theories. We begin 
by studying the branes of the superaffine theory. We are interested in the perturbation
by the bulk field 
\beq
\Phi \eq J^3  \, \bar{J}^3 \eq  4 \, \partial_{z} X \, \bar\partial_{\bar{z}} X \ ,
\eeq
where the factor of $4$ takes into account that the currents are differently
normalised at $k=2$. 
As we have explained above in section~3, $J^3$ corresponds to the 
current $\partial X$ of the superaffine theory that survives the orbifold
from the circle theory. Thus $\Phi$ describes indeed the radius-changing
modulus we are interested in. Note also that this field is the 
$G_{-1/2} \bar{G}_{-1/2}$ descendant of the $h=\bar{h}=\tfrac{1}{2}$ 
field $\chi\bar\chi$ and hence preserves the ${\cal N}=1$ superconformal
symmetry in the bulk.

\subsection{The superaffine case}

In terms of the WZW model, the analysis is essentially identical to what
was done in \cite{PERT} and reviewed in section~2. In fact, the level of
the WZW model only enters in a rather trivial way, namely as an overall
factor in front of (\ref{bulkboundary}), and hence the calculation and the
conclusions are exactly as described there. In terms of the 
superaffine boundary states, this then implies that  
\begin{align}
| \! | \left( \begin{smallmatrix} 0 &ie^{i\psi} \\ ie^{-i\psi} &0 \end{smallmatrix} 
\right);\eta \rangle \! \rangle_{sa} \ \
\quad  \stackrel{\de R < 0}{\Longleftarrow} \quad \ \ 
 | \! | g;\eta \rangle \! \rangle_{sa} \ \ 
 \quad \stackrel{\de R > 0}{\Longrightarrow} \quad
 \ \ | \! | \left( \begin{smallmatrix} e^{i\phi} &0 \\ 0 &e^{-i\phi} 
 \end{smallmatrix} \right);\eta \rangle \! \rangle_{sa} \ .
\label{5,9}
\end{align}
As was explained in section~4.1, the diagonal and off-diagonal boundary conditions 
correspond to Dirichlet and Neumann branes, whose position and 
Wilson line is determined by the phase of the unperturbed SU(2) element. 
Thus we conclude that the superaffine branes flow to (a superposition of)
Dirichlet or Neumann branes as the radius is increased or decreased,
respectively.\footnote{From the point of view of the superaffine theory,
all of these branes are however fundamental and cannot be further
resolved.} This mirrors precisely the result obtained in \cite{PERT} for the
bosonic $c=1$ theory. This conclusion is independent of whether these branes are 
BPS or non-BPS.

\subsection{The circle theory at $R=1$}
\label{sec:circleflow}

Having understood the boundary flow for the superaffine theory,
we can now use the fact that the circle theory at radius $R=1$
is the ${\mathbb Z}_2$ orbifold of the superaffine theory, to 
deduce what happens in the circle theory. The quantum symmetry
by means of which we can obtain the circle theory from the superaffine
theory was already given in (\ref{quo}). By construction, this orbifold 
projects out the twisted sector of the original ${\cal S}$-orbifold; 
in particular, it removes all states of half-integer winding from the 
spectrum (see (\ref{2,3})). On the superaffine branes, the orbifold acts as 
$\tilde{\cal S}\, | \! | g;\eta \rangle \! \rangle_{sa} = 
| \! | \si_{3}g \si_{3};\eta \rangle \! \rangle_{sa}$;  the orbifold invariant
boundary states (that define the boundary states of the circle theory) are then
\cite{MP}
\beq
| \! | g;\eta \rangle \! \rangle_{sc} \ = \ 
\frac{1}{\sqrt{2}} \Bigl( | \! | g;\eta \rangle \! \rangle_{sa} \ 
+ \ | \! | \si_{3} \, g \, \si_{3} ;\eta \rangle \! \rangle_{sa} \Bigr) \ .
\label{6,2}
\eeq
Note that the brane associated to $g$ is identical to the one associated to 
$\si_{3}g\si_{3}$, and thus the resulting brane moduli space is 
SU(2)/$\ZZ_{2}$ where the $\ZZ_{2}$ changes the sign of the 
off-diagonal entries of $g$.\footnote{It is therefore different from the  
$g \equiv -g$ equivalence in SO(3).} Expressed in terms
of the parameters of (\ref{1,F}), conjugation by 
$\si_{3}$ simply corresponds to the shift $\psi\mapsto \psi+\pi$, but
does not affect $\theta$ (nor $\phi$). The radius perturbation, on the other
hand, only affects $\theta$, and thus the RG flow is compatible with the
$\tilde{\cal S}$ orbifold. Alternatively, we can think about how the disc
correlation functions (from which the bulk boundary coefficient that appears
in the RG equation can be deduced) behave under the orbifold: since
the currents in question live in a sector that is invariant under the orbifold
action, the result is unchanged, and thus the old analysis applies. 
We can therefore conclude that 
\begin{align}
| \! | \left( \begin{smallmatrix} 0 &ie^{i\psi} \\ ie^{-i\psi} &0 
\end{smallmatrix} \right);\eta \rangle \! \rangle_{sc} \ \ 
\quad \stackrel{\de R < 0}{\Longleftarrow} \quad
\ \ | \! | g;\eta \rangle \! \rangle_{sc} \ \
\quad \stackrel{\de R > 0}{\Longrightarrow} \quad
\ \ | \! | \left( \begin{smallmatrix} e^{i\phi} &0 \\ 0 &e^{-i\phi} \end{smallmatrix} 
\right);\eta \rangle \! \rangle_{sc} \ .
\label{6,3}
\end{align}
Thus, as before, the superconformal branes of the circle theory
flow to Dirichlet or Neumann branes as the radius is increased or
decreased, respectively. There is however now a new subtlety: 
the end-point of the RG flow is not necessarily a fundamental
brane in the circle theory. For example, for $\eta=-$, the
resulting diagonal group element (to which the system flows if the
radius is increased) describes a superposition of a BPS
Dirichlet brane and anti-brane at opposite points on the circle, while for
$\eta=+$ the diagonal group element describes a superposition of
two non-BPS Dirichlet branes at opposite points on the circle. 
Unlike the situation in the superaffine orbifold, these branes are 
not fundamental in the circle theory.

\subsection{The circle theory at $R=\frac{M}{N}$}
\label{sec:CircleMNflow}

Finally, we want to comment on the situation where the radius of the 
superconformal circle theory is rational. As suggested in \cite{PERT}
(see also \cite{Tseng:2002ax}) 
in the context of the bosonic analysis, we can make use of the fact 
that the circle theory at radius $R=\frac{M}{N}$ can be obtained as a 
$\ZZ_{M} \times \ZZ_{N}$ orbifold of the circle theory at radius $R=1$.
In fact, the relevant orbifold action can be taken to be 
\beq
{\cal S}_{N} \ := \ \Bigl( X \ \ \mapsto \ \ X \ + \ \frac{2\pi R}{N} \Bigr) \co 
{\cal W}_{M} \ := \ \Bigl( \tilde{X} \ \ \mapsto \ \ \tilde{X} \ + \ \frac{\pi}{RM} \Bigr) \ .
\label{6,4}
\eeq
The branes of the fractional radius theory can then be obtained from the
branes at $R=1$ in the usual manner. The only subtlety involves the
determination of the fixed points. One finds that fixed points only appear 
if $N$ is even: the non-BPS branes ($\eta=+$) of the $R=1$ theory are 
fixed points under ${\cal S}_{N}^{N/2}$. The resolution of these fixed points
then leads to the inclusion of a R-R component in the boundary state, and thus
the branes with $\eta=+$ are BPS for $N$ even. (On the other hand, the 
R-R part of the BPS branes at $R=1$ is projected out for $N$ even, and hence
the $\eta=-$ branes are non-BPS.) Thus for even $N$ the roles of the BPS and 
non-BPS branes are interchanged, in perfect agreement with what was already found
(by some different reasoning) in \cite{GA}. 

Just like the $\tilde{\cal S}$ orbifold action in section~5.2, the orbifold operators 
${\cal S}_{N}$ and ${\cal W}_{M}$ act on the boundary states as 
$g \mapsto \Ga_{M} g \Ga_{M}^{-1}$ and $g \mapsto \Ga_{N} g \Ga_{N}$, 
respectively, where $\Ga_L=\te{diag}(e^{\frac{i\pi}{L}},e^{-\frac{i\pi}{L}})$ is 
defined as in \cite{GA}. In particular, these operators therefore only act on the 
phases $\phi$, $\psi$ in (\ref{1,F}), but not 
on the modulus angle $\theta$. (They also leave the sector in which the 
perturbing field lives invariant.) Thus as before the radius changing orbifold
does not affect the RG flow analysis, and the result goes through directly. In
general, though, the end-point of the RG flow will now be a superposition
of a number of Dirichlet or Neumann branes. 

\section{Conclusions}
\label{sec:Conclusions}

In this paper we have studied the behaviour of the ${\cal N}=1$
superconformal boundary conditions of the free boson and free
fermion theory at $c=\tfrac{3}{2}$ under the radius-changing bulk
deformation. The results are similar to those that were previously 
obtained in the bosonic case in \cite{PERT}: if the radius is increased,
a generic brane flows to a superposition of Dirichlet branes, while
the endpoint of a radius decreasing perturbation is a superposition
of Neumann branes.

As in the bosonic example, our analysis hinged on relating the
circle theory to an SU(2) WZW model for which the RG flow can
be solved explicitly. In the present context, the WZW model in
question appears at level $k=2$, and it is equivalent to 
the superaffine theory, rather than the circle theory directly. 
In addition, there were some subtleties involving the GSO projection.

\section*{Acknowledgements}

This research has been partially supported by 
the Swiss National Science Foundation and the Marie Curie network
`Constituents, Fundamental Forces and Symmetries of the Universe'
(MRTN-CT-2004-005104). This paper is largely based on the 
Diploma thesis of O.S.

\appendix
\renewcommand{\theequation}{A.\arabic{equation}}

\section{Comparison of overlaps}
\label{sec:appA}
\setcounter{equation}{0}

In order to identify the boundary states of the superaffine
theory with the WZW model it is useful to compare their
overlaps. We begin with the analysis of the superaffine
boundary states given in (\ref{3,1}). Using the same
techniques as in \cite{GA} one finds that their overlap
equals
\begin{align}
_{sa} \langle \! \langle g_{1} ; - | \! | \, q^{L_{0} - \frac{c}{24}} \,
 | \! | g_{2}; - \rangle \! \rangle_{sa} \ & \; = \
 \frac{1}{2} 
  \sum_{n \in \ZZ} 
\Bigl(f_{3}(\tilde{q}) \ + \ (-1)^{n} \, f_{4}(\tilde{q})\Bigr) \, 
 \frac{\tilde{q}^{\frac{1}{2} \left( -\frac{\al}{\pi} + n \right)^{2}}}{\eta(\tilde{q})} \notag \\
_{sa} \langle \! \langle {g}_{1} ; + | \! | \, q^{L_{0} - \frac{c}{24}} \,
 | \! | {g}_{2}; + \rangle \! \rangle_{sa} \ & \; = \
  f_{3}(\tilde{q}) \sum_{n \in \ZZ}   \frac{\tilde{q}^{\frac{1}{2} 
  \left( -\frac{\al}{\pi} + n \right)^{2}}}{\eta(\tilde{q})} \notag \ ,
\end{align}
where $\tilde{q}$ is the variable in the open string channel, and 
\beq 
\cos (\al ) \ = \ \frac{1}{2}\, \te{Tr} \Bigl( g_{1}^{-1} \, g_{2} \Bigr) \ .
\label{A,1}
\eeq
Here we have used the standard $f_i$ functions from \cite{Pol} that are 
defined as 
\begin{eqnarray}
&& f_2(q) \ =   \ \sqrt{2} \, q^{\frac{1}{24}} \prod_{n=1}^{\infty} (1+q^n)  \qquad
f_3(q) \ =  \ q^{-\frac{1}{48}} \prod_{n=1}^{\infty} (1+q^{n+\frac{1}{2}}) \nonumber \\
&& \qquad \qquad \qquad \qquad 
 f_4(q) \ = \ q^{-\frac{1}{48}} \prod_{n=1}^{\infty} (1-q^{n+\frac{1}{2}}) \ . 
\label{ffun}
\end{eqnarray}
The calculation of the overlap between boundary states corresponding to different
values of $\eta$ depends on our convention concerning the relative normalisation 
of the NS-NS Ishibashi states; the convention we have used is that \footnote{This differs from the conventions of \cite{MP} by a factor of $(-1)^{m}$.}
 \beq
^{\te{NS}}\langle \! \langle j;m,n;\eta | \, q^{L_{0} - \frac{c}{24}} \, 
|j;m,n;-\eta \rangle \! \rangle^{\te{NS}} \ = \  
(-1)^{j} \, f_{4}(q) \; 
\frac{q^{\frac{j^{2}}{2}} \ + \ q^{\frac{(j+1)^{2}}{2}} }{\eta(q)} \ .
\label{A,2}
\eeq
Then one finds that 
\beq
_{sa} \langle \! \langle g_{1} ; - | \! | \, q^{L_{0} - \frac{c}{24}} \, 
| \! | {g}_{2}; + \rangle \! \rangle_{sa} \ = \ 
\frac{f_{2}(\tilde{q})}{\sqrt{2}}  \sum_{n \in \ZZ-\frac{1}{2}} 
\frac{\tilde{q}^{\frac{1}{2} \left( -\frac{\al}{\pi} + n \right)^{2}}}{\eta(\tilde{q})} \ .
\label{A,3}
\eeq
To compare to the usual level 2 WZW characters $\chi_{j}$, we take $g_1=\pm g_2$. Then one obtains
\begin{align}
_{sa} \langle \! \langle g ; - | \! | \, q^{L_{0} - \frac{c}{24}} \, | \! |  g; - \rangle \! \rangle_{sa} \ \ 
&=  \ \frac{1}{2}  \sum_{n \in \ZZ} \Bigl(f_{3}(\tilde{q}) + \ (-1)^{n} \, 
f_{4}(\tilde{q})\Bigr) \;  \frac{\tilde{q}^{\frac{1}{2} n^{2}}}{\eta(\tilde{q})}
 \eq \chi_{j=0}(\tilde{q}) \notag \\
 _{sa} \langle \! \langle g ; - | \! | \, q^{L_{0} - \frac{c}{24}} \, | \! | \!-\! g; - \rangle \! \rangle_{sa} \ \ 
&= \  \frac{1}{2} \sum_{n \in \ZZ} \Bigl(f_{3}(\tilde{q}) - \ (-1)^{n} \, 
f_{4}(\tilde{q})\Bigr) \;  \frac{\tilde{q}^{\frac{1}{2} n^{2}}}{\eta(\tilde{q})}
 \eq \chi_{j=1}(\tilde{q}) \notag \\
_{sa} \langle \! \langle g ; + | \! | \, q^{L_{0} - \frac{c}{24}} \,
 | \! | g ; + \rangle \! \rangle_{sa} \ &= \ \sum_{n \in \ZZ} 
 f_{3}(\tilde{q}) \; \frac{\tilde{q}^{\frac{1}{2}  n^{2}}}{\eta(\tilde{q})} \eq 
 \chi_{j=0}(\tilde{q}) \ + \ \chi_{j=1}(\tilde{q}) \notag \\
_{sa} \langle \! \langle g ; - | \! | \, q^{L_{0} - \frac{c}{24}} \,
 | \! | g; + \rangle \! \rangle_{sa} \  &= \ \sum_{n \in \ZZ-\frac{1}{2}} 
 \frac{f_{2}(\tilde{q})}{\sqrt{2}} \; \frac{\tilde{q}^{\frac{1}{2}  n^{2}}}{\eta(\tilde{q})} 
 \eq \chi_{j=\frac{1}{2}}(\tilde{q}) \ ,
\label{3,21}
\end{align}
where we have used standard theta-function identities (see for example
\cite{GR}) to relate the expressions
to the characters of the WZW model given in (\ref{ffch}). The expressions on the
right-hand side precisely agree with what one expects, based on the fusion rules
of the su(2) level $2$ theory. It is also not difficult to see how both sides generalise
for general $g_1$ and $g_2$.

\end{document}